\documentclass{iau}
\usepackage{graphicx}
\usepackage[OT4]{fontenc}

\usepackage{url}

\def\Mpc{$h^{-1}$~{\rm  Mpc}}

\medskip

\title[High redshift QSO systems] %% give here short title %%
{Tracing high redshift cosmic web with quasar systems}

\author[Maret Einasto]   %% give here short author list %%
{Maret Einasto$^1$}

\affiliation{$^1$Tartu Observatory, \\ Observatooriumi 1, 61602
  T\~oravere, Estonia  \\ email: {\tt maret.einasto@to.ee} }

\pubyear{2015}
\volume{308}  %% insert here IAU Symposium No.
\jname{The Zeldovich Universe: Genesis and Growth of the Cosmic Web}
\editors{R. van de Weygaert, S. Shandarin, E. Saar \&  J. Einasto, eds.}
\begin{document}
\maketitle

\begin{abstract}
We study the cosmic web at redshifts  $1.0 \leq z \leq 1.8$
using quasar systems based on quasar data from the SDSS DR7 QSO catalogue.
Quasar systems were determined with a friend-of-friend (FoF) algorithm 
at a series of linking lengths. 
At the linking lengths $l \leq 30$~\Mpc\ the diameters of quasar systems are
smaller than the diameters of random systems, and are
comparable to the sizes of galaxy superclusters in the local Universe.
The  mean space density of quasar systems is
close to the mean space density of local rich superclusters.
At larger linking lengths the diameters of quasar systems are comparable with
the sizes of supercluster complexes in our cosmic neighbourhood. 
The richest quasar systems have diameters exceeding $500$~\Mpc.
Very rich systems can be found also in random distribution but
the percolating system which penetrate the whole sample volume
appears in quasar sample at smaller linking
length than in random samples showing that the large-scale distribution
of quasar systems differs from random distribution.
Quasar system catalogues at our web pages
(\url{http://www.aai.ee/~maret/QSOsystems.html})
serve as a database to search for
superclusters of galaxies and to trace the
cosmic web at high redshifts.

\end{abstract}

\firstsection 
\section{Introduction}
According to the contemporary cosmological paradigm the 
cosmic web formed and evolved  from  tiny density
perturbations in the very early Universe by hierarchical
growth driven by  gravity (\cite{2009LNP...665..291V} and 
references therein).
To understand  how the  cosmic web formed and  evolved we need
to describe and quantify it at  low and  high redshifts.  
Large galaxy
redshift  surveys  like  SDSS  enable  us to  describe  the  cosmic  web  in our
neighbourhood in detail. 
One source of information about  the cosmic structures  at
high redshifts is the distribution of quasars --- energetic nuclei of massive galaxies.
Already decades ago several studies described large systems
in quasar distribution 
(\cite{1982MNRAS.199..683W}, \cite{1991MNRAS.249..218C}, 
\cite{2012MNRAS.419..556C}, \cite{2013MNRAS.429.2910C})
which are known as Large Quasar Groups (LQGs). 
LQGs may trace  distant galaxy superclusters (\cite{1996MNRAS.282..713K}). 
The large-scale distribution of quasar systems
gives us information about the cosmic web at high redshifts
which are not yet covered by large and wide galaxy surveys. 

The aim of our study is to study the high redshift
cosmic web using data about quasar systems. We find quasar systems
and analyse their properties and large-scale distribution at
redshifts $1.0 \leq z \leq 1.8$ using quasar data
from \cite{2010AJ....139.2360S} catalogue of quasars,
based on the Sloan Digital Sky Survey Data Release 7. 

We select from this catalogue a subsample of quasars in the redshift
interval $1.0 \leq z \leq 1.8$, and apply $i$-magnitude limit
$i = 19.1$. In order to reduce the edge effects of our analysis, we limit the
data in the area of SDSS sky coordinate limits $-55  \leq \lambda  \leq 55$
degrees and $-33  \leq \eta  \leq 35$ degrees. 
Our final sample contains data of 22381 quasars.
The mean space density of quasars is very low, approximately
$1.1 \cdot 10^{-6}\mathrm{(h^{-1}Mpc)}^{-3}$, therefore it is important 
to understand whether their distribution
differs from random distribution. To compare quasar and
random distributions we generated random samples with the same number
of points, and sky coordinate and redshift limits as quasar samples. 

We assume  the standard cosmological parameters: the Hubble parameter $H_0=100~ 
h$ km~s$^{-1}$ Mpc$^{-1}$, the matter density $\Omega_{\rm m} = 0.27$, and the 
dark energy density $\Omega_{\Lambda} = 0.73$.

\section{Results}

We determined quasar and random systems with the friend-of-friend (FoF) 
algorithm at a series of linking lengths and
present catalogues of quasar systems. 
FoF method collects objects into systems if they have at least one 
common neighbour closer than a linking length. 
At each linking length we found the number of systems
in quasar and random samples with at least two members,
calculated multiplicity functions of
systems, and analysed the richness and size of systems.
For details we refer to \cite{2014A&A...5568A..46E}.

\begin{figure}[ht]
\centering
\resizebox{0.45\textwidth}{!}{\includegraphics*{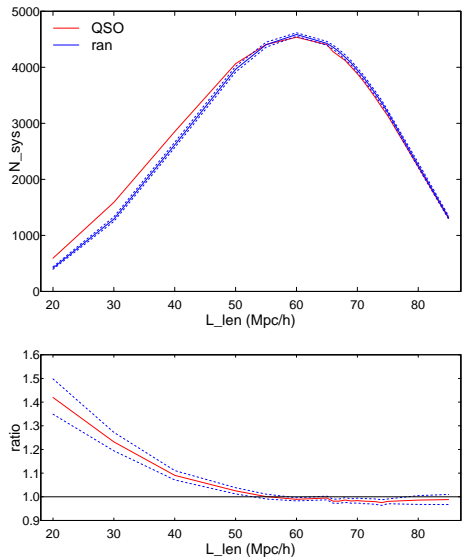}}\\
\caption{The number of quasar and random systems (upper panel) and the
ratio of the numbers of quasar and random systems (lower panel)
vs. the linking length. 
Red solid line denote quasar systems, and blue  lines denote random
systems, black line shows ratio 1.
}
\label{fig:nsysll}
\end{figure}

\begin{figure}[ht]
\centering
\resizebox{0.45\textwidth}{!}{\includegraphics*{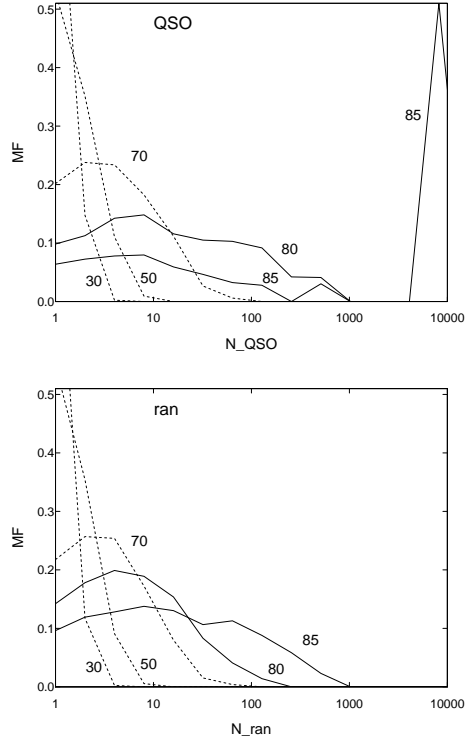}}\\
\caption{Multiplicity functions MF (the fraction of 
systems of different richness) of quasar (upper panel) and random (lower panel)
data at the linking lengths $30$, $50$, $70$, $80$,
and $85$~\Mpc.
}
\label{fig:mf}
\end{figure}

Up to the linking lengths approximately $50$~\Mpc\ 
the number of quasar systems is larger than the number of systems in random 
catalogues (Fig.~\ref{fig:nsysll}), at larger linking
lengths the number of systems becomes similar to that in random catalogue.
The number of systems in both quasar and random catalogues
reaches maximum at $60$~\Mpc. At higher values of the linking lengths
systems begin to join into larger systems  and the number of systems 
decreases. Multiplicity functions in Fig.~\ref{fig:mf} show that at the linking
length $l = 85$~\Mpc\ about half of quasars join the richest quasar system ---
a percolation occurs.  In random catalogues the richest system is much smaller
than the richest quasar system. Therefore FoF analysis shows that the
distribution of quasars and the properties 
of quasar systems differ from random  at small and large linking lengths.

\begin{figure}[ht]
\centering
\resizebox{0.40\textwidth}{!}{\includegraphics*{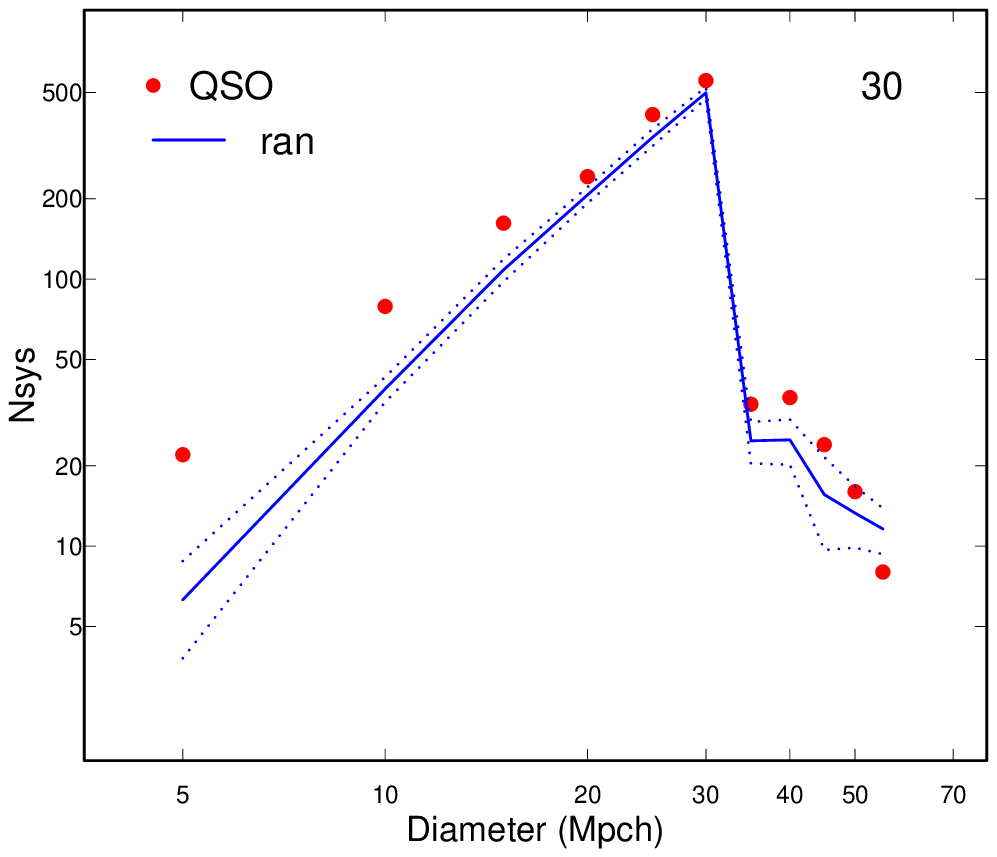}}
\resizebox{0.40\textwidth}{!}{\includegraphics*{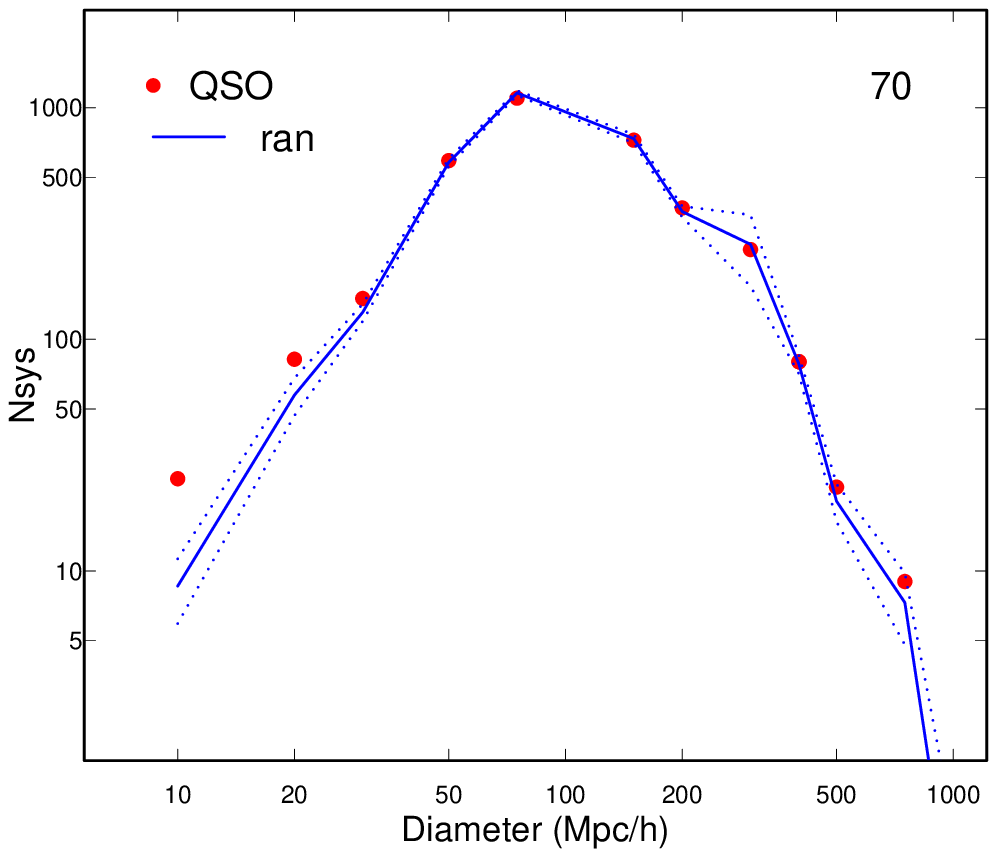}}\\
\caption{Number of quasar (red dots) and random (blue lines)
systems of different diameter for linking lengths $30$ and $70$~\Mpc\
(upper and lower panel, correspondingly).
}
\label{fig:diam2070}
\end{figure}

In Fig.~\ref{fig:diam2070} we compare the distribution of diameters 
(maximum distance between quasar pairs in a system)
of quasar and random systems at the  linking lengths $30$ and $70$~\Mpc. 
At the linking length $30$~\Mpc\ in the whole diameter interval 
the number of quasar systems with a given diameter
is higher than that of random systems, the difference is statistically highly
significant. At larger linking lengths ($l \geq 40$~\Mpc; we show this for $70$~\Mpc) 
the number of quasar
systems with diameters up to $20$~\Mpc\ is always larger than the number of
random systems at these diameters. From diameters 
$\approx 30$~\Mpc\ the number of systems of different diameter
in quasar and random catalogues becomes similar. Among both quasar and random
systems there are several very large systems
with diameters larger than $500$~\Mpc. 

\begin{figure}[ht]
\centering
\resizebox{0.45\textwidth}{!}{\includegraphics*{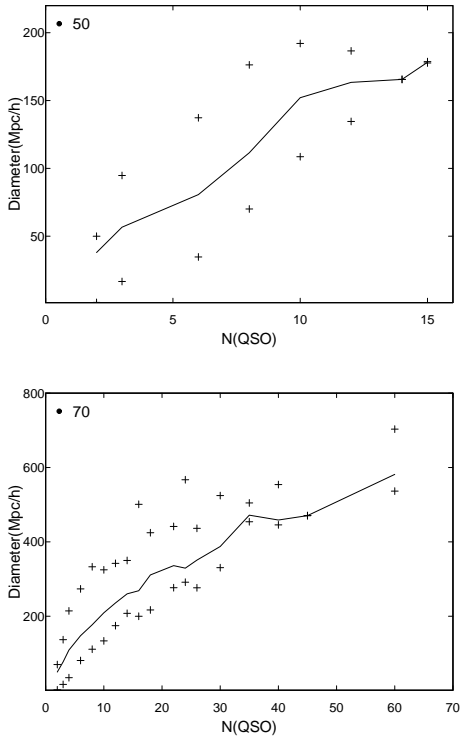}}\\
\caption{System richness $N_{QSO}$ vs. their diameter $D_{max}$ for 
quasars for linking lengths $50$ and $70$~\Mpc.
Lines show median values of diameters, crosses denote 
the smallest and the largest diameters.
}
\label{fig:qso70nsysdmax}
\end{figure}

In Fig.~\ref{fig:qso70nsysdmax} we show the median, 
and minimum and maximum values of quasar system 
diameters vs. their richness at the linking lengths $50$ and $70$~\Mpc.
At the linking length $50$~\Mpc\ the sizes of the richest
quasar systems, $\approx 200$~\Mpc, are comparable to the sizes 
of the richest superclusters in the local Universe
(\cite{1994MNRAS.269..301E}). The mean space density of quasar systems 
of order of $10^{-7}\mathrm{(h^{-1}Mpc)}^{-3}$, this is
close to the mean space density of local rich superclusters
(\cite{1997A&AS..123..119E}).

The sizes of the largest quasar systems at $l = 70$~\Mpc, $500 - 700$~\Mpc,
are comparable with the sizes of supercluster complexes 
in the local Universe (\cite{2011ApJ...736...51E}, \cite{2012A&A...539A..80L}).  
At this linking length we obtain systems of the same size also from the random
catalogues. 

\begin{figure}[ht]
\centering
\resizebox{0.70\textwidth}{!}{\includegraphics*{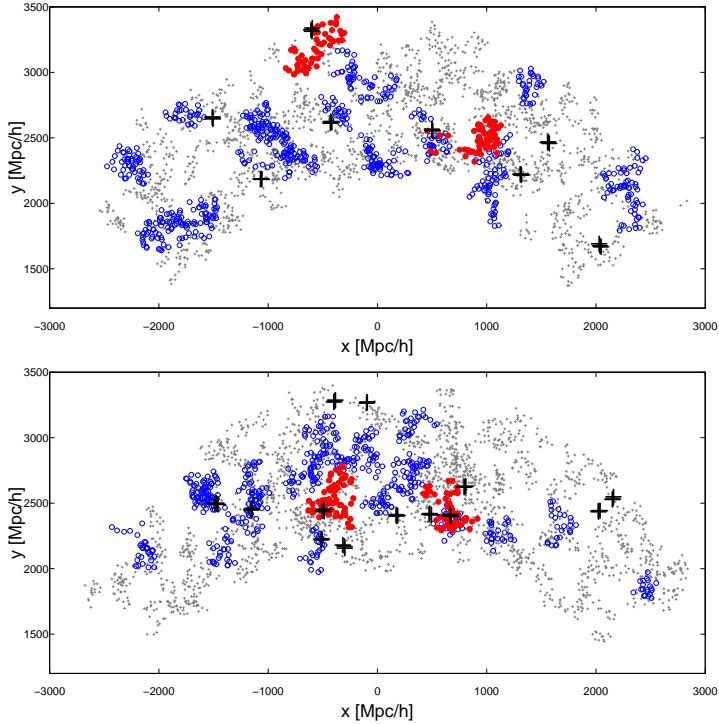}}\\
\caption{Distribution of QSO systems at the linking length $70$~\Mpc\ in $x$ and $y$ coordinates
in two slices by $z$ coordinate (upper panel:
$z \leq 0$~\Mpc, lower panel: $z > 0$~\Mpc).
Grey dots denote quasars in systems with $10 \leq N_{QSO} \leq 24$,
blue circles denote quasars in systems with $25 \leq N_{QSO} \leq 49$,
and red filled circles denote quasars in systems with $N_{QSO} \geq 50$.
Black crosses denote quasar triplets at  a linking length $20$~\Mpc.
}
\label{fig:qso70xy}
\end{figure}

We show in Fig.~\ref{fig:qso70xy} the distribution 
of quasars in systems of various richness at linking length $70$~\Mpc\ in 
cartesian coordinates $x$, $y$, and $z$ 
(see \cite{2014A&A...5568A..46E}):
\begin{equation}
\begin{array}{l}
    x = -d \sin\lambda, \nonumber\\[3pt]
    y = d \cos\lambda \cos \eta,\\[3pt]
    z = d \cos\lambda \sin \eta,\nonumber
\end{array}
\label{eq:xyz}
\end{equation}
where $d$ is the comoving distance, and $\lambda$ and $\eta$ are the SDSS 
survey coordinates. 
We plot in the figure also quasars from the richest systems at 
$l = 20$~\Mpc\ (quasar triplets).

Visual inspection of Fig.~\ref{fig:qso70xy} shows that very rich quasar systems
form a certain pattern. In some areas of the figure
there are underdense regions between rich quasar systems
with diameters of about $400$~\Mpc\ (e.q. in the upper panel
between $-1000 < x < 1000$~\Mpc). 
The size of underdense regions in this figure is much larger than
the sizes of typical large voids in the local Universe
(see \cite{2011A&A...534A.128E})
but is close to the sizes of the largest voids 
covered by SDSS survey (\cite{2011A&A...532A...5E}, \cite{2012ApJ...759L...7P}).
Very rich systems were found also
from random catalogues but the percolation analysis shows that the large-scale distribution
of quasar systems differs from random distribution. 
We shall analyse the large scale distribution of quasar systems
in detail in another study.

The richest system at the linking length 
$l = 70$~\Mpc\ at $x \approx 1000$~\Mpc\ and $y \approx 2500$~\Mpc\
is the Huge-LQG described in \cite{2013MNRAS.429.2910C}.
The presence of very rich systems as supercluster complexes is an essential
property of the cosmic web, and do not violate homogeneity of the universe at very large 
scales, as claimed by \cite{2013MNRAS.429.2910C}.

\section{Summary}

We determined quasar systems at a series of linking lengths, and found
that at small linking lengths their diameters and space density are similar
to those of rich galaxy superclusters in the local Universe. 
At the linking lengths $l \geq 50$~\Mpc\ the diameters of the 
richest quasar systems are comparable with
the sizes of supercluster complexes in our cosmic neighbourhood, 
exceeding $500$~\Mpc. Systems of similar richness were determined also in
random catalogues but the large-scale distribution
of quasar systems differs from random distribution. 
We may conclude that quasar systems as markers of galaxy superclusters
and supercluster complexes give us a snapshot of the high-redshift cosmic web. 
Quasar system catalogues serve as a database to search for
high-redshift superclusters of galaxies and to trace the
cosmic web at high redshifts.

I thank my coauthors Erik Tago, Heidi Lietzen,  Changbom Park, Pekka Hein\"am\"aki,
Enn Saar, Hyunmi Song, Lauri Juhan Liivam\"agi Jaan Einasto
for enjoyable and fruitful collaboration.

The present study was supported by ETAG project 
IUT26-2, and by the European Structural Funds
grant for the Centre of Excellence "Dark Matter in (Astro)particle Physics and
Cosmology" TK120.


\begin{thebibliography}{55}
\expandafter\ifx\csname natexlab\endcsname\relax\def\natexlab#1{#1}\fi

\bibitem[Clowes \& Campusano(1991)]{1991MNRAS.249..218C}
{Clowes}, R.~G. \& {Campusano}, L.~E. 1991, \textit{MNRAS}, 249, 218

\bibitem[{Clowes} {et~al.}(2012)]{2012MNRAS.419..556C}
{Clowes}, R.~G., {Campusano}, L.~E., {Graham}, M.~J., \& {S{\"o}chting}, I.~K.
  2012, \textit{MNRAS}, 419, 556

\bibitem[{Clowes} {et~al.}(2013)]{2013MNRAS.429.2910C}
{Clowes}, R.~G., {Harris}, K.~A., {Raghunathan}, S., {et~al.} 2013, \textit{MNRAS},
  429, 2910

\bibitem[{Einasto} {et~al.} 1994]{1994MNRAS.269..301E}
{Einasto}, M., {Einasto}, J., {Tago}, E., {Dalton}, G.~B., \& {Andernach}, H.
  1994, \textit{MNRAS}, 269, 301
  
\bibitem[{Einasto} {et~al.} 1997]{1997A&AS..123..119E}
{Einasto}, M., {Tago}, E., {Jaaniste}, J., {Einasto}, J., \& {Andernach}, H.
  1997, \textit{A\&AS}, 123, 119

\bibitem[{Einasto} {et~al.}(2011c)]{2011ApJ...736...51E}
{Einasto}, M., {Liivam{\"a}gi}, L.~J., {Tempel}, E., {et~al.}
  2011{\natexlab{c}}, \textit{ApJ}, 736, 51
  
\bibitem[{Einasto} {et~al.}(2011b)]{2011A&A...532A...5E}
{Einasto}, M., {Liivam{\"a}gi}, L.~J., {Tago}, E., {et~al.} 2011{\natexlab{b}},
  \textit{A\&A}, 532, A5

\bibitem[{Einasto} {et~al.}(2011a)]{2011A&A...534A.128E}
{Einasto}, J., {Suhhonenko}, I., {H{\"u}tsi}, G., {et~al.} 2011{\natexlab{a}},
  \textit{A\&A}, 534, A128
  

\bibitem[{Einasto} {et~al.}(2014)]{2014A&A...5568A..46E}
{Einasto}, M. and {Tago}, E. and {Lietzen}, H.  {et~al.} 2014,
  \textit{A\&A}, 568, A46

\bibitem[{Komberg} {et~al.} 1996]{1996MNRAS.282..713K}
{Komberg}, B.~V., {Kravtsov}, A.~V., \& {Lukash}, V.~N. 1996, \textit{MNRAS}, 282, 713

\bibitem[{Liivam{\"a}gi} {et~al.}(2012)]{2012A&A...539A..80L}
{Liivam{\"a}gi}, L.~J., {Tempel}, E., \& {Saar}, E. 2012, \textit{A\&A}, 539, A80

\bibitem[{Park} {et~al.}(2012)]{2012ApJ...759L...7P}
{Park}, C., {Choi}, Y.-Y., {Kim}, J., {et~al.} 2012, \textit{ApJL}, 759, L7
  
  
\bibitem[{Schneider} {et~al.}(2010)]{2010AJ....139.2360S}
{Schneider}, D.~P., {Richards}, G.~T., {Hall}, P.~B., {et~al.} 2010, \textit{AJ}, 139,
  2360

\bibitem[{van de Weygaert} \& {Schaap} (2009)]{2009LNP...665..291V}
{van de Weygaert}, R. \& {Schaap}, W. 2009, in Lecture Notes in Physics, Berlin
  Springer Verlag, Vol. 665, Data Analysis in Cosmology, ed. V.~J.
  {Mart{\'{\i}}nez}, E.~{Saar}, E.~{Mart{\'{\i}}nez-Gonz{\'a}lez}, \& M.-J.
  {Pons-Border{\'{\i}}a}, 291--413

\bibitem[{Webster}(1982)]{1982MNRAS.199..683W}
{Webster}, A. 1982, \textit{MNRAS}, 199, 683



\end{thebibliography}
\end{document}